\documentclass{SPAICE}
\usepackage{xcolor}
\def\authorEmail{carlo.cena@polito.it}
\def\AuthorShort{A. Ayanmanesh Motlaghmofrad et al.}

\author[1]{Amirhossein Ayanmanesh Motlaghmofrad}
\author[1,2]{Carlo Cena\thanks{Corresponding author. E-Mail: \authorEmail}\thanks{This work was supported by the the project PNRR-NGEU which has received funding from the MUR – DM 117/2023.}}
\author[1]{Mauro Martini}
\author[1]{Marcello Chiaberge}
\affil[1]{Politecnico di Torino, Torino (TO), ITA}
\affil[2]{Argotec S.R.L., San Mauro (TO), ITA}

\title{MPC for underactuated spacecraft control with a Lyapunov supervised physics-informed neural network correction layer}

\begin{document}

% Creates the title and author list automatically for you!
\maketitle

\begin{abstract}
Underactuated spacecraft faces controllability limitations and heightened sensitivity to environmental disturbances, complicating attitude maneuvering and stabilization. Due to the lack of control authority along the underactuated axis, conventional controllers cannot directly stabilize all attitude components and therefore require reference planning strategies. Furthermore, MPC approaches remain sensitive to inertia uncertainty and unmodeled dynamic couplings, resulting in degraded tracking performance under mismatch. To address these issues, we consider a hierarchical architecture integrating three layers: (i) a nonlinear model predictive controller (NMPC) for constraint and underactuation-aware maneuver planning and nominal closed-loop stability under actuator limits; (ii) a physics-informed neural network (PINN) trained offline on simulation data to estimate residual disturbance torques, with loss terms that enforce consistency with rigid-body rotational dynamics; (iii) a Lyapunov-based supervisory safety mechanism that evaluates the learned correction online and bounds or suppresses its influence to preserve the stability properties of the baseline controller. The architecture is evaluated in a high-fidelity simulation environment modelling reaction wheel dynamics, actuator saturation, and environmental disturbances. Experimental studies show statistically significant reductions in steady-state attitude error relative to standalone NMPC while maintaining robust behavior under uncertainty. The supervisory layer ensures graceful degradation to purely model-based control when the learning-based augmentation is unreliable.
\end{abstract}

% EDIT HERE
% The main document. Please add your content as desired.
% We provide examples for adding figures, equations, and tables. Please stick to the style used in this template.
%\textbf{Every paper has to contain an abstract, an introduction section, a results section and a discussion section.} You can add intermediate and/or subsections as required. \textbf{Please do not change the style of this template}.
% references to tables, figures and sections with \cref
%For multi-line equations, use align
\section{Introduction}
Attitude control is fundamental to spacecraft operations, enabling precise maneuvers, Earth and deep-space observation, and reliable communication between space and ground. However, design decisions or actuator failures can result in underactuated spacecraft, reducing control over rotational dynamics and necessitating careful characterization of the actuation system. Common actuators include control moment gyros (CMGs), reaction wheels (RWs), and thrusters. When only internal torques are available, control is constrained by the conservation of angular momentum, leading to fundamental accessibility limitations \cite{1103519}. For the study of underactuated spacecraft, two principal approaches have been proposed. The first assumes zero total angular momentum, which simplifies the dynamics and enables short-time local controllability of the reduced attitude system, allowing rest-to-rest maneuvers between admissible configurations \cite{att_stab}. The second exploits predictable external torques, such as solar radiation pressure, to recover full three-axis attitude control without relying on conservation assumptions \cite{aerospace9090498, react}. With a practical approach in mind, the present study adopts the zero-total-angular-momentum assumption. Although environmental disturbances may induce drift in the otherwise uncontrollable component of angular momentum over time, it is assumed that occasional external momentum management counteracts this drift. Recent advancements in space technology have explored neural networks for trajectory optimization and adaptive feedback control, suitable for onboard implementation \cite{SHIROBOKOV202187, davide}. Applications include adaptive neural-network controllers \cite{KRISHNAKUMAR1995131}, neural networks assisting primary controllers \cite{APOLLONI1997279}, and neural-network-based optimal attitude control with impulsive thrusters \cite{nn_opt_att_imp}. Physics-informed approaches, such as physics-informed normalizing flows, have also been used in several applications \cite{pinn_rizzo, pinn_loss_eletr, cena2026learningrobustsatelliteattitude, pinn_sat_state_est}. Additionally, in the broader control field data-driven MPC approaches are increasingly explored \cite{annurev_datadriven_mpc, davide}. This work proposes a hybrid model-based and learning-based control architecture that systematically combines the strengths of both paradigms while mitigating their standalone limitations. A model-based baseline controller ensures nominal stability, for the controllable axes, and safety properties, forming a certifiable backbone suitable for safety-critical operations. This baseline is augmented with a physics-informed neural network (PINN) operating as a residual torque compensator, enabling improved tracking performance in the presence of underactuation and structured modeling uncertainties. To preserve closed-loop guarantees, a deterministic Lyapunov-based supervisory mechanism constrains the neural network contribution in real time, ensuring that stability and safety properties remain intact in accordance with formal guidance \cite{ecss_mlh}.

The main contributions of this paper can be summarized as follows:
\begin{figure*}[ht]
    \centering
    \includegraphics[width=0.97\linewidth]{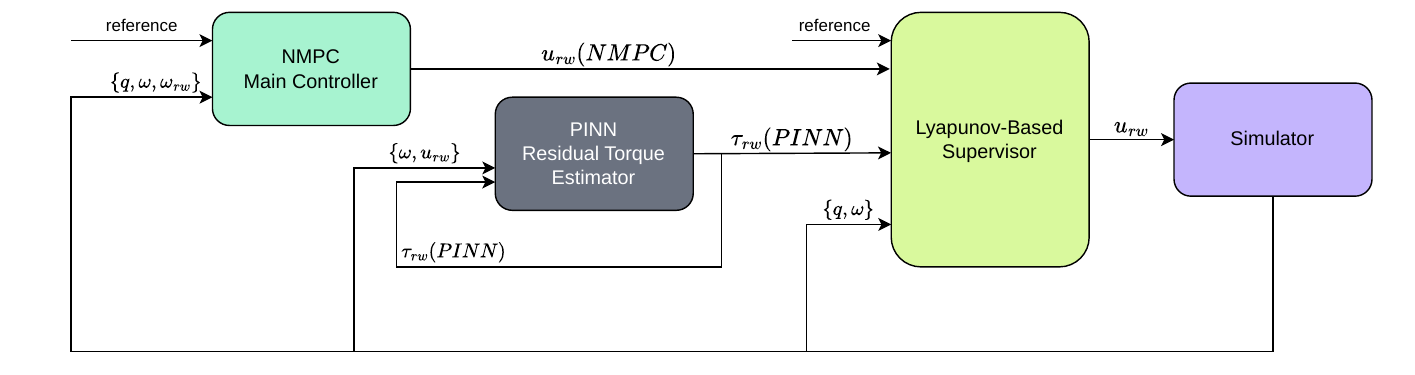}
	\caption{High-level block diagram representation of the proposed control system architecture for the underactuated spacecraft under study.}
	\label{fig:block_diag}
\end{figure*}

\begin{itemize}
    \item We propose a novel hybrid control framework that combines a baseline model-based controller with a physics-informed neural network (PINN) to address underactuation and modeling uncertainties, while preserving theoretical guarantees for safety-critical operations.
    \item We develop a deterministic Lyapunov-based supervisory mechanism that constrains the neural network contribution in real time, in accordance with formal verification principles \cite{ecss_mlh}.
    \item We show improved steady-state tracking performance, achieving a 3.8\% reduction in average steady-state error and a 24.16\% reduction for the worst case compared to the baseline controller, and demonstrate significant improvements in the final error, yielding a 12.7\% reduction in average final error, 52.9\% reduction in the maximum final error, and 29.42\% reduction in final error variance.
\end{itemize}

\section{Methodology}
\label{sec:methodology}
The control architecture is hierarchical and comprises three main components: a nonlinear model predictive controller (NMPC) as the baseline policy, a physics-informed neural network (PINN) that estimates residual disturbance torques to reduce model-plant mismatch, and a Lyapunov-based supervisory safety layer that ensures stability and constraint satisfaction. The attitude is represented by a unit quaternion \( q \) and body rates \( \omega \). The reaction-wheel speeds are denoted by \( \omega_{\mathrm{RW}} \), and the input is the wheel motor torque \( u_{\mathrm{RW}} \). Practical limits are treated as hard constraints, including wheel speed \( |\omega_{\mathrm{RW}}| \le 6000 \, \mathrm{rpm} \), motor torque \( \|u_{\mathrm{RW}}\|_\infty \le 0.05 \, \mathrm{Nm} \), stored momentum \( \approx 0.9 \, \mathrm{Nms} \), and torque slew-rate \( \|\Delta u_{\mathrm{RW}}\|_\infty \le \Delta u_{\max} \) with \( \dot{u}_{\max} = 0.4 \, \mathrm{Nm/s} \) and \( T_s = 0.5 \, \mathrm{s} \). %The plant model uses rigid-body kinematics with quaternion renormalization and coupled spacecraft-wheel rotational dynamics.
The nominal predictive model assumes a diagonal inertia \( J \) with underactuated actuator distribution and discretization by fixed-step RK4. Gravity-gradient and aerodynamic drag are included, while solar radiation pressure and magnetic torques are neglected as secondary in the considered regime. The control objective is to achieve a steady-state pointing error bounded by a \(0.5^\circ\) settling-angle requirement, satisfy all actuator and operational constraints at all times, and preserve robust closed-loop behavior under bounded model mismatch. The NMPC formulation employs a discrete prediction model from RK4 with an update period \( T_{\mathrm{NMPC}} = 0.5 \, \mathrm{s} \) and horizon length \( N_h = 24 \) (12s window). The optimal control problem minimizes a finite-horizon cost with stage terms penalizing attitude error, angular-rate error, control effort, and input increments, plus a quadratic terminal cost. Whereas, the PINN learns a bounded residual torque \( \tau_d \) as a corrective term in the augmented discrete-time dynamics of \cref{eq:discr_dyn}.
\begin{equation}
\label{eq:discr_dyn}
    x_{k+1} = \Phi_{\mathrm{RK4}}(x_k, u_{\mathrm{RW},k}, \tau_d(k)),
\end{equation}
where \( \Phi_{\mathrm{RK4}}(\cdot) \) denotes the fourth-order Runge--Kutta discretization of the continuous dynamics. The residual torque is assumed constant over each NMPC sampling interval, \( \tau_d(t) \approx \tau_d(k) \) for \( t \in [kT_s, (k+1)T_s) \), reflecting the fact that dominant model mismatch sources, such as inertial parameter errors and slowly varying disturbances, evolve on time scales significantly slower than the NMPC update rate. The network follows a NARX structure with three hidden layers of 128 neurons each with tanh activations. It receives as input 16 features: the previous 2 estimated disturbance torque, and a set of 14 engineered features created starting from the current and previous spacecraft angular velocity, and the current RWs torques. Inputs are normalized using training set statistics, and outputs are scaled by conservative bounds to enforce magnitude limits. The training objective is formulated as a multi-step rollout loss through the dynamics. Starting from state \( x_k \), the augmented dynamics are propagated over \( H = 10 \) integration steps (sampled ten times faster than the NMPC update frequency), yielding a predicted angular velocity \( \hat{\omega}_{k+H}(\theta) \). The physics-informed loss is defined in \cref{eq:loss_pi}.
\begin{equation}
    \label{eq:loss_pi}
    \mathcal{L}(\theta) = \sum_{k \in \mathcal{K}} \| \hat{\omega}_{k+H}(\theta) - \omega_{k+H} \|_2^2,
\end{equation}
where \( \omega_{k+H} \) denotes the measured angular velocity from simulation data, \( \mathcal{K} \) is the set of training time indices, and \( \theta \) represents the network parameters. By construction, the loss depends on \( \theta \) only through the effect of \( \tau_d(k) \) on state evolution over the prediction horizon. This formulation aligns the learning objective with the intended role of the PINN: improving short-horizon prediction accuracy in a manner consistent with its deployment in the closed-loop system. The Lyapunov-based supervisory layer does not stabilize the spacecraft independently; its role is to avoid that the learned correction degrades the closed-loop stability already achieved by the nominal NMPC. At each control update, the supervisor evaluates whether injecting the PINN correction is compatible with a non-increasing Lyapunov-like energy associated with the attitude error dynamics; if this condition is violated, the correction is attenuated or rejected while the nominal NMPC command is left unchanged.

Let \( q_e = q^{-1}\otimes q_{\mathrm{ref}} \) and \( \omega_e = \omega - \omega_{\mathrm{ref}} \) denote the attitude and rate tracking errors. The candidate Lyapunov function is
\begin{equation}
    V(q_e,\omega_e) = \tfrac{1}{2}\omega_e^\top J \omega_e + k_q\,(1-q_{e,0}), \qquad k_q>0,
\end{equation}
positive definite about \( q_e=[1,0,0,0]^\top,\ \omega_e=0 \). For the rest-to-rest objective considered here (\( \dot\omega_{\mathrm{ref}}\approx 0 \)), its derivative along the rigid-body dynamics \( J\dot\omega = \tau - \omega\times(J\omega) \) is
\begin{equation}
    \dot V = \omega_e^\top\big(\tau - \omega\times(J\omega)\big) - \frac{k_q}{2} q_{e,0}\, q_{e,v}^\top \omega_e .
\end{equation}
The applied torque is decomposed as \( \tau = \tau_{\mathrm{nom}} + \tau_{\mathrm{corr}} \), where \( \tau_{\mathrm{nom}} = -B_{\mathrm{act}} u_{\mathrm{NMPC}} \) is generated by the NMPC and \( \tau_{\mathrm{corr}} = -B_{\mathrm{act}} u_{\mathrm{PINN}} \) is the realization of the PINN's disturbance estimate \( \tau_d \). Since an ideal correction satisfies \( \tau_{\mathrm{corr}} = -\tau_d \), a learning failure with a sign-flipped estimate would double rather than cancel the disturbance term in \cref{eq:discr_dyn}. To guard against this failure mode, the supervisor tests the Lyapunov condition under a conservative assumption: it introduces a scalar attenuation factor \( \alpha\in[0,1] \) and evaluates \( \dot V \) under the test torque
\begin{equation}
    \tau_{\mathrm{test}} = \tau_{\mathrm{nom}} - 2\alpha\,\tau_{\mathrm{corr}},
\end{equation}
selecting the largest \( \alpha \) for which \( \dot V(\tau_{\mathrm{test}}) \le 0 \); if no \( \alpha>0 \) satisfies this condition, the correction is rejected (\( \alpha=0 \)) and the closed loop reverts to the pure NMPC command. This condition is enforced as a local safeguard. Additional conservative bounds are applied at implementation level: the PINN correction torque is saturated to \( 0.015\,\mathrm{Nm} \), the total commanded torque remains within the \( 0.05\,\mathrm{Nm} \) actuator limit, and the correction is disabled whenever the attitude error falls below \( 1^\circ \). Overall, the closed loop consists of the NMPC baseline performing both implicit trajectory planning and feedback control, to generate constraint-aware command references that actively reduce the attitude error along the underactuated axis, while ensuring nominal stability and recursive feasibility; the PINN providing a bounded, physics-regularized residual torque estimate that drives the spacecraft's behavior closer to the NMPC's predictive model; and the Lyapunov supervisor enforcing a safety cage that guarantees graceful degradation to pure NMPC when learning is unreliable. 

\section{Experimental Setup}
\label{sec:exp_setup}
The validation campaign quantitatively evaluates whether PINN-based residual torque compensation improves attitude regulation performance relative to a baseline NMPC controller. All experiments are conducted under controlled and repeatable simulation conditions with fixed controller parameters established prior to testing. We performed a testing campaign spanning diverse initial attitudes, by generating $50$ unit direction vectors uniformly distributed on the unit sphere $\mathbb{S}^2$ using a Fibonacci lattice sampling scheme. For each direction, two initial attitudes are defined by applying rotations of $75^\circ$ and $125^\circ$ about the corresponding axis, yielding $100$ distinct initial attitude quaternions, to ensure comprehensive coverage of both moderate and large-angle slew maneuvers. The spacecraft and all reaction wheels are initialized with zero angular velocity, establishing rest-to-rest conditions with zero total angular momentum. All simulations incorporate representative environmental disturbances for low Earth orbit operations, including gravity-gradient torques and aerodynamic drag effects. Model mismatch is deliberately introduced through uncertainty in the spacecraft inertia tensor: the principal moments of inertia in the prediction model are perturbed with respect to the true values used in simulation, introducing structured parametric uncertainty in both diagonal and off-diagonal components. This structured model error creates prediction discrepancies that motivate residual learning compensation. Disturbance and uncertainty realizations remain fixed across paired simulations to isolate the effects of control architecture differences. Each simulation lasts $100$ seconds, providing sufficient duration to observe both transient response and steady-state behavior. Convergence is declared when the attitude error magnitude remains below $0.5^\circ$ for the remainder of the trajectory.

\textbf{Controller Configurations:}
Three controllers are compared under identical conditions, including reference profiles, actuator constraints, and numerical integration settings, without controller retuning. The tested controllers are:
\begin{enumerate}
    \item \textit{Baseline NMPC}: Standard nonlinear model predictive control using the nominal prediction model without learning augmentation.
    \item \textit{NMPC + PINN}: NMPC augmented with PINN-based residual torque correction applied directly to the control input.
    \item \textit{NMPC + PINN + Supervisor}: PINN-augmented NMPC with residual corrections filtered through a Lyapunov-based supervisory layer prior to actuation.
\end{enumerate}
\begin{table*}[ht!]
\begin{center}
\begin{tabular}{cccc} 
\toprule
\textbf{Metric [deg]} & \textit{NMPC} & \begin{tabular}{@{}c@{}}\textit{NMPC}\\ \textit{PINN}\end{tabular} & \begin{tabular}{@{}c@{}}\textit{NMPC}\\ \textit{PINN+Sup}\end{tabular}\\
\midrule
SS RMSE avg         & $0.1476\pm0.0672$ & $\textbf{0.1385}\pm\textbf{0.0437}$ & $0.1419\pm0.0495$ \\
SS RMSE max         & $0.4940$ & $\textbf{0.3392}$ & $0.3747$ \\
Final error avg     & $0.0870\pm0.0704$ & $\textbf{0.0722}\pm\textbf{0.0452}$ & $0.0759\pm0.0497$ \\
Final error max     & $0.4915$ & $\textbf{0.1607}$ & $0.2315$ \\
\bottomrule
\end{tabular}
\caption{Results obtained with $100$ simulations using the three controller introduced in \cref{sec:exp_setup}.}
\label{tab:results}
\end{center}
\end{table*}
\textbf{Metrics:}
Performance is quantified using two attitude-error-based metrics computed for each test realization:
\begin{itemize}
    \item Steady-state RMSE: Root mean square attitude error computed over all time samples satisfying the $0.5^\circ$ convergence criterion, characterizing steady-state precision.
    \item Final attitude error: Attitude error magnitude at the terminal simulation time, quantifying the final pointing accuracy.
\end{itemize}
For each controller, we aggregate metrics across all test runs, reporting mean values and standard deviations to characterize dispersion across initial conditions. Statistical significance is assessed through paired comparisons between each learning-augmented configuration and its corresponding baseline realization using the Wilcoxon signed-rank test. This nonparametric approach avoids normality assumptions and evaluates whether paired differences are systematically shifted from zero, thereby distinguishing consistent performance improvements from incidental gains on isolated trajectories. Additionally, we show the closed-loop time history of the \textit{NMPC + PINN + Supervisor} controller to provide complementary qualitative insights into transient behavior and convergence characteristics.

\section{Results}
\label{sec:results}
\begin{figure}[t]
    \centering
    \includegraphics[width=0.93\columnwidth]{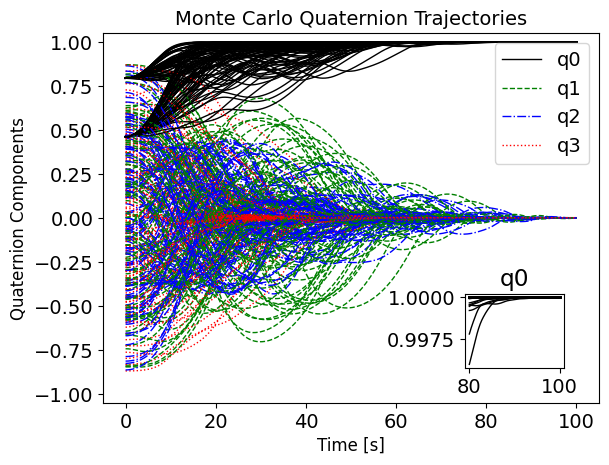}
	\caption{Test quaternion trajectories under NMPC with PINN-based residual compensation and Lyapunov-based supervisory filtering.}
	\label{fig:mc_res}
\end{figure}

%Aggregate Test Performance
Aggregate statistical metrics computed over the full testing campaign are shown in \cref{tab:results}. Mean steady-state RMSE values remain well below the prescribed $0.5^\circ$ accuracy bound for all cases, indicating that observed variations occur within the high-precision regulation regime required by the task.
% \begin{table}[t]
% \begin{center}
% \begin{tabular}{ccc} 
% \toprule
% \textbf{\begin{tabular}{@{}c@{}}\textbf{Metric}\\ \textbf{improvement [\%]}\end{tabular}} & \begin{tabular}{@{}c@{}}\textit{NMPC}\\ \textit{PINN}\end{tabular} & \begin{tabular}{@{}c@{}}\textit{NMPC}\\ \textit{PINN+Sup}\end{tabular}\\
% \midrule
% SS RMSE avg         & $\textbf{6.13}$ & $3.82$ \\
% SS RMSE max         & $\textbf{31.35}$ & $24.16$ \\
% SS RMSE std         & $\textbf{35.01}$ & $26.32$ \\
% Final error avg     & $\textbf{16.93}$ & $12.74$ \\
% Final error max     & $\textbf{67.30}$ & $52.90$ \\
% Final error std     & $\textbf{35.77}$ & $29.42$ \\
% \bottomrule
% \end{tabular}
% \caption{Percentage improvements relative to the NMPC baseline for the results shown in \cref{tab:results}.}
% \label{tab:results_impr}
% \end{center}
% \end{table}
%Statistical Significance Analysis
Additionally, paired comparisons with respect to the baseline NMPC controller were performed using the Wilcoxon signed-rank test. Both learning-augmented configurations demonstrate statistically significant improvements relative to the baseline. The results support the hypothesis that PINN-based residual torque estimation enhances NMPC performance. The learned compensation systematically reduces both steady state RMSE and final attitude error across the test runs; a behavior that is consistent with the role of the PINN in correcting structured model mismatch, such as inertia uncertainty and unmodeled dynamic coupling, thereby causing the plant to more closely match the NMPC's internal predictive model. Moreover, improvements are not limited to typical trajectories: both the steady-state and final error dispersions decrease under learning-augmented control. This reduction in variance and improvement in terms of worst case (maximum error in \cref{tab:results}), together with statistically significant paired improvements, demonstrates that the PINN enhances not only the nominal accuracy but also the uniformity and reliability of performance across diverse initial attitudes. While the supervised PINN configuration outperforms the baseline controller, its improvements are smaller than those of the unsupervised variant \textit{NMPC + PINN}. Though, this outcome is expected, as the supervisory layer enforces a conservative Lyapunov-based non-increase condition using the nominal inertia model. As a result, beneficial corrections may be partially rejected when they appear unsafe under the nominal assumptions. The supervisor thus prioritizes stability over performance, yielding a controlled trade off: constrained use of the learned correction in exchange for preserving the guarantees of the baseline NMPC controller. Finally, \cref{fig:mc_res} shows the closed-loop attitude trajectories obtained under the hierarchical controller, which exhibit consistent convergence to the reference equilibrium.

\section{Discussion}
\label{sec:discussion}
This work presented a hierarchical control framework for attitude regulation of underactuated spacecraft that integrates NMPC, PINN-based residual compensation, and deterministic safety supervision. The nominal NMPC controller provides the primary control authority, handling nonlinear dynamics, actuator limits, and underactuation-aware maneuver planning. A physics‑informed neural network is added as a residual torque estimator to compensate for model mismatch arising primarily from inertia uncertainty and neglected coupling effects. Finally, a Lyapunov-based supervisory layer ensures safety by filtering the learned correction through an energy‑based admissibility condition. If the residual estimate is unreliable or rejected by the supervisor, the closed-loop system automatically reverts to the nominal NMPC policy without mode switching. This design aligns with safety‑critical requirements typical of aerospace applications. The architecture was implemented in a high‑fidelity simulation environment, accounting for real‑time execution constraints, scheduling at multiple frequencies, numerical robustness, and consistency between training assumptions and closed‑loop operating conditions. A test validation campaign showed that PINN‑based residual compensation improves steady‑state attitude accuracy and reduces performance dispersion. When used without supervision, the largest performance gains were observed; with the supervisory layer active, improvements remained statistically significant while preserving strict stability guarantees, highlighting a clear performance-safety trade‑off. A relevant aspect concerns the computational and architectural complexity introduced by the proposed method relative to the observed performance gain. The PINN adds a feedforward evaluation at each NMPC sampling interval that does not affect the real-time feasibility of the $T_s = 0.5\,\mathrm{s}$ control loop, as shown in \cite{cena2026learningrobustsatelliteattitude}. The Lyapunov supervisor further adds only a closed-form scalar admissibility check, incurring no additional optimization overhead. While the average improvement in steady-state RMSE is modest, the architecture yields larger gains in the worst-case and dispersion metrics (24.16\% reduction in maximum steady-state RMSE, 29.42\% reduction in final error variance). Given the low computational cost of the added components, we argue that even a modest average improvement is justified when accompanied by a substantial reduction in worst-case error.

\begin{acknowledgments}
    This work has been developed with the contribution of the Politecnico di Torino Interdepartmental Centre for Service Robotics (PIC4SeR https://pic4ser.polito.it).
\end{acknowledgments}

\printbibliography

@misc{ecss_mlh,
  author       = {{European Cooperation for Space Standardization (ECSS)}},
  title        = {{ECSS-E-HB-40-02A: Space engineering - Machine learning handbook. Handbook}},
  year         = {2024}
}

@article{SHIROBOKOV202187,
title = {Survey of machine learning techniques in spacecraft control design},
journal = {Acta Astronautica},
volume = {186},
pages = {87-97},
year = {2021},
issn = {0094-5765},
doi = {10.1016/j.actaastro.2021.05.018},
author = {Maksim Shirobokov and Sergey Trofimov and Mikhail Ovchinnikov}
}

@INPROCEEDINGS{pinn_sat_state_est,
  author={Varey, Jacob and Ruprecht, Jessica D. and Tierney, Michael and Sullenberger, Ryan},
  booktitle={2024 IEEE Aerospace Conference}, 
  title={Physics-Informed Neural Networks for Satellite State Estimation}, 
  year={2024},
  volume={},
  number={},
  pages={1-8},
  doi={10.1109/AERO58975.2024.10521414}}

@article{davide,
  author={Celestini, Davide and Gammelli, Daniele and Guffanti, Tommaso and D'Amico, Simone and Capello, Elisa and Pavone, Marco},
  journal={IEEE Robotics and Automation Letters}, 
  title={Transformer-Based Model Predictive Control: Trajectory Optimization via Sequence Modeling}, 
  year={2024},
  volume={9},
  number={11},
  pages={9820-9827},
  doi={10.1109/LRA.2024.3466069}}

@ARTICLE{pinn_loss_eletr,
  author={Falas, Solon and Asprou, Markos and Konstntinou, Charalambos and Michael, Maria K.},
  journal={IEEE Transactions on Industrial Informatics}, 
  title={{Robust Power System State Estimation Using Physics-Informed Neural Networks}}, 
  year={2025},
  volume={pre-print}}

@article{pinn_rizzo,
title = {Physics-informed Neural Network for Quadrotor Dynamical Modeling},
journal = {Robotics and Autonomous Systems},
volume = {171},
pages = {104569},
year = {2024},
issn = {0921-8890},
doi = {10.1016/j.robot.2023.104569},
author = {Weibin Gu and Stefano Primatesta and Alessandro Rizzo},
}

@Article{aerospace9090498,
AUTHOR = {Jin, Lei and Li, Yingjie},
TITLE = {Model Predictive Control-Based Attitude Control of Under-Actuated Spacecraft Using Solar Radiation Pressure},
JOURNAL = {Aerospace},
VOLUME = {9},
YEAR = {2022},
NUMBER = {9},
ARTICLE-NUMBER = {498},
ISSN = {2226-4310},
DOI = {10.3390/aerospace9090498}
}

@ARTICLE{1103519,
  author={Crouch, P.},
  journal={IEEE Transactions on Automatic Control}, 
  title={Spacecraft attitude control and stabilization: Applications of geometric control theory to rigid body models}, 
  year={1984},
  volume={29},
  number={4},
  pages={321-331},
  doi={10.1109/TAC.1984.1103519}}

@article{att_stab,
author = {Krishnan, Hariharan and McClamroch, N. Harris and Reyhanoglu, Mahmut},
title = {Attitude stabilization of a rigid spacecraft using two momentum wheel actuators},
journal = {Journal of Guidance, Control, and Dynamics},
volume = {18},
number = {2},
pages = {256-263},
year = {1995},
doi = {10.2514/3.21378}
}

@article{cena2026learningrobustsatelliteattitude,
    title = {Learning robust satellite attitude dynamics with physics-informed normalizing flow},
    journal = {Acta Astronautica},
    volume = {245},
    pages = {970-981},
    year = {2026},
    doi = {https://doi.org/10.1016/j.actaastro.2026.03.047},
    author = {Carlo Cena and Mauro Martini and Marcello Chiaberge},
}

@article{annurev_datadriven_mpc,
   author = "Berberich, Julian and Allgöwer, Frank",
   title = "An Overview of Systems-Theoretic Guarantees in Data-Driven Model Predictive Control", 
   journal= "Annual Review of Control, Robotics, and Autonomous Systems",
   year = "2025",
   volume = "8",
   number = "Volume 8, 2025",
   pages = "77-100",
   doi = "10.1146/annurev-control-030323-024328",
   publisher = "Annual Reviews",
   issn = "2573-5144",
  }

@article{APOLLONI1997279,
title = {A co-operating neural approach for spacecrafts attitude control},
journal = {Neurocomputing},
volume = {16},
number = {4},
pages = {279-307},
year = {1997},
issn = {0925-2312},
doi = {10.1016/S0925-2312(97)00035-0},
author = {B. Apolloni and F. Battini and C. Lucisano}
}

@inproceedings{react,
  author       = {Cena, Carlo and Bucci, Silvia and Balossino, Alessandro},
  title        = {{Deep reinforcement learning for under-actuated satellite attitude control and reaction wheel desaturation using solar radiation pressure}},
  booktitle    = {Proceedings of the International Astronautical Congress (IAC)},
  month        = {10},
  year         = {2023}
}

@article{KRISHNAKUMAR1995131,
title = {Adaptive neuro-control for spacecraft attitude control},
journal = {Neurocomputing},
volume = {9},
number = {2},
pages = {131-148},
year = {1995},
note = {Control and Robotics, Part II},
issn = {0925-2312},
doi = {10.1016/0925-2312(94)00062-W},
author = {K. KrishnaKumar and S. Rickard and S. Bartholomew}
}

@article{nn_opt_att_imp,
author = {Biggs, James D. and Fournier, Hugo},
title = {Neural-Network-Based Optimal Attitude Control Using Four Impulsive Thrusters},
journal = {Journal of Guidance, Control, and Dynamics},
volume = {43},
number = {2},
pages = {299-309},
year = {2020},
doi = {10.2514/1.G004226}
}
\addcontentsline{toc}{section}{References}

\end{document}